\documentclass{JHEP3} 

\usepackage{epsfig,multicol}
\usepackage{amsmath, amssymb}
\usepackage{concmath, palatino}

\newcommand\fverb{\setbox\pippobox=\hbox\bgroup\verb}
\newcommand\fverbdo{\egroup\medskip\noindent%
			\fbox{\unhbox\pippobox}\ }
\newcommand\fverbit{\egroup\item[\fbox{\unhbox\pippobox}]}
\newbox\pippobox

\newcommand{\be}{\begin{equation}}
\newcommand{\ee}{\end{equation}}
\newcommand{\ba}{\begin{eqnarray}}
\newcommand{\ea}{\end{eqnarray}}

\newcommand{\pint}{\makebox[0pt][l]{\hspace{2.4pt}$-$}\int}
\newcommand{\oh}[1]{{\cal O}( #1 )}

\newcommand{\refeq}[1]{Eq.~(\ref{eq:#1})}
\newcommand{\reffig}[1]{Fig.~(\ref{fig:#1})}

\newcommand{\la}{\longrightarrow}

\newcommand{\ads}{AdS_5\times S^5}

\title{The generalized scaling function of AdS/CFT and semiclassical string theory}

\author{Matteo Beccaria\\
  Dipartimento di Fisica, Universita' del Salento,
  Via Arnesano, 73100 Lecce\\
  INFN, Sezione di Lecce\\
  E-mail: \email{matteo.beccaria@le.infn.it}}

\abstract{
Recently, Freyhult, Rej and Staudacher (FRS) proposed an integral equation determining the leading logarithmic term 
of the anomalous dimension of $\mathfrak{sl}(2)$ twist-operators in ${\cal N}=4$ SYM for large Lorentz spin $M$ and 
twist $L$ at fixed $j = L/\log\,M$. We discuss the large $j$ limit of the FRS equation.
This limit can be matched with the {\em fast long string} limit of $\ads$ superstring perturbation 
theory at all couplings.  
In particular, a certain part of the classical and one-loop string result is known to be protected and can be computed in 
the weakly coupled large-$j$ limit of the FRS equation.
We present various analytical and numerical results supporting agreement at one and two loops 
in the gauge theory. 
}

\begin{document}

\section{Introduction}

The anomalous dimensions of Wilson twist operators~\cite{Belitsky:2003ys} are relevant perturbative quantities
which appear in various phenomenological problems in the study of QCD strong interactions. A typical example
is the operator product expansion analysis of deep inelastic scattering~\cite{Predazzi}.
In that context, the close relation between parton splitting functions and anomalous dimensions suggests various 
physical insights valid in special kinematical limits. In particular, the behavior of anomalous dimensions 
for large Lorentz spin $M$ at fixed twist $L$ probes the quasi-elastic limit where the Bjorken variable is close to unity
$x_{\rm Bj}\to 1$.
In this regime the most singular part of the splitting functions is due to soft gluon emission and is universal. 
For the leading twist 2 operators, these remarks translate into the following well-known prediction for the 
anomalous dimension $\gamma$
\be
\label{eq:logscaling}
\gamma = 2\,\Gamma_{\rm cusp}(g)\,\log\,M  + \oh{M^0},
\ee
where $g^2 = \frac{\lambda}{16\,\pi^2}$ and $\lambda = N_c\,g^2_{\rm YM}$ is  
the 't Hooft planar coupling $\lambda$. The non trivial function $\Gamma_{\rm cusp}(g)$ is the so-called 
cusp anomalous dimension~\cite{cusp}.

The logarithmic scaling in \refeq{logscaling} is quite general and applies in particular to the superconformal finite 
${\cal N}=4$ SYM theory where integrability~\cite{Beisert:2005fw} and 
AdS/CFT duality~\cite{Maldacena:1997re,Gubser:1998bc,Witten:1998qj} can be exploited 
to gain (much) additional information. This approach is clearly interesting in itself due to the theoretical
relevance of the ${\cal N}=4$ SYM theory. Besides, one can also argue that large $x_{\rm Bj}$ physics 
can be related to the QCD one, being mostly related to the shared gauge sector. A recent example of this strategy
is the analysis of a generalized Gribov-Lipatov 
reciprocity~\cite{Basso:2006nk,Dokshitzer:2006nm,Dokshitzer:2005bf}
for various twist-2 and twist-3 Wilson 
operators~\cite{Beccaria:2007cn,Beccaria:2007vh,Beccaria:2007bb,Beccaria:2007pb,Beccaria:2008fi}.

The current knowledge of $\Gamma_{\rm cusp}(g)$ in ${\cal N}=4$ SYM is quite complete. It can be extracted
from the anomalous dimensions of $\mathfrak{sl}(2)$ operators. The weak-coupling 
perturbative series can be computed at all-orders by a rather simple expansion of the so-called BES equation~\cite{Beisert:2006ez}.
The result is in agreement with the most advanced available field theoretical computations~\cite{cusp2}.
The problem of computing the strong coupling expansion of the BES equation is more difficult and after intense activity~\cite{cuspstrong}
has been impressively solved in the remarkable paper~\cite{Basso:2007wd}. Again, there is full agreement with the 
two-loop analysis of the dual superstring theory on $\ads$~\cite{Gubser:2002tv,Frolov:2002av,Roiban:2007jf,Roiban:2007dq}.

The BES equation is derived by considering operators with an arbitrary finite twist $L$ and taking the large spin limit $M\to\infty$. If the twist increases
with the spin $M$ then one expects a richer landscape of scaling behaviors. A simple one-loop illustration of this general statement can be found 
in~\cite{Belitsky:2006en}. It is shown that when $M\gg L$, one must still distinguish between two quite different  regimes
characterized by extreme values of the gauge theory parameter $\xi$ defined as
\be
\xi = \frac{1}{L}\,\log\frac{M}{L}.
\ee
In particular, the {\em minimal} anomalous dimension has the following leading contributions
\ba
\label{eq:BMNone}
\gamma(g, M) = \left\{\begin{array}{ll}
\displaystyle 8\,g^2\,\log{M}, & \qquad \xi \gg 1 , \\ \\
\displaystyle 8\,g^2\,\frac{1}{L}\,\log^2\frac{M}{L}, & \qquad \xi \ll 1.
\end{array}
\right.
\ea
The first case is covered by the BES equation. The second case with the characteristic double logarithm enhancement is beyond its reach.
The appearance of these two regimes is in quite similarity with the semiclassical string calculation of ~\cite{Frolov:2002av}
as we shell discuss in a moment.

In~\cite{Freyhult:2007pz}, Freyhult, Rej and Staudacher (FRS) proposed to analyze the logarithmic behavior
of anomalous dimensions in the following limit 
\be
\label{eq:limit}
L, M\to \infty,\qquad j = \frac{L}{\log\,M} = \mbox{fixed}.
\ee
In this limit FRS prove that a logarithmic scaling is observed once more. The prefactor now depends on 
both $g$ and $j$
\be
\gamma(g, j) = f(g, j)\,\log\,M + \oh{M^0},
\ee
where $f(g,j)$ is a generalization of the cusp anomalous dimension
\be
f(g, 0) \equiv f(g) = 2\,\Gamma_{\rm cusp}(g).
\ee
An integral equation analogous to the BES equation, but valid for all $g$ and $j$, 
has been derived in~\cite{Freyhult:2007pz}.  Of course, a great deal of interesting results can be obtained 
by applying to the FRS equation the methods which have been already sharpened in the case $j=0$. 
In particular, this means that the FRS equation can be considered in the  following two opposite limits.

\medskip
\noindent\underline{\em 1. The fully weak limit}. 
This is simply $g,j\to 0$. 
There seems to be no ambiguity in this double limit
and it is convenient to first expand $f(g, j)$ around $j=0$
\be
f(g, j) = \sum_{n\ge 0} f_n(g)\,j^n,
\ee
and then expand each $f_n(g)$ around $g=0$
\be
f_n(g) = \sum_{k\ge 0} {\cal F}_{n,k}\,g^{2\,k}.
\ee
The coefficients ${\cal F}_{n,k}$ have been computed in~\cite{Freyhult:2007pz} where their explicit expression can be found as well
as a discussion of various features, like for instance transcendentality uniformity.

\medskip
\noindent\underline{\em 2. The Alday-Maldacena limit}.
The Alday-Maldacena (AM) limit is a strong coupling limit defined by the general condition $g\to\infty$ with $j\ll g$~\cite{Alday:2007mf}. The 
scaling function $f(g,j)$ is described in this limit by the thermodynamical Bethe Ansatz equations of the 
non linear $O(6)$ $\sigma$-model~\cite{sigmamodel}. As explained in the beautiful analysis of~\cite{Basso:2008tx} it is necessary to consider 
separately the
two situations where $j\ll m$ or $j\gg m$ where $m$ is the dynamically generated mass gap~\cite{massgap}
\be
m = \frac{2^{3/4}\,\pi^{1/4}}{\Gamma(5/4)}\,g^{1/4}\,e^{-\pi\,g}\,\left(1+{\cal O}(1/g)\right).
\ee 
In particular, the case $j\ll m \ll g$ predicts the large $g$ behavior of the functions $f_n(g)$ and can be summarized
by the expansion
\be
\label{eq:sigma}
f(g,j) = -j+m^2\,\left[\frac{j}{m}+\frac{\pi^2}{24}\,\left(\frac{j}{m}\right)^3+\cdots\right],
\ee
which has been indeed recovered in the FRS equation in~\cite{Basso:2008tx}.
For additional numerical and analytical confirmations of the expansion \refeq{sigma} see~\cite{Fioravanti:2008rv,Fioravanti:2008ak}. 
Additional terms in the above series which represent the $\sigma$-model energy density can be found in~\cite{Buccheri:2008ap}.
The other limit $m\ll j \ll g$ is also very interesting and is discussed in details in~\cite{Basso:2008tx}.

\medskip
The above two limits are similar to those already considered for the cusp anomalous dimension since 
the parameter $j$ is used as a perturbative book-keeping device. This suggests to consider another new limit.

\medskip
\noindent\underline{\em 3. The large-$j$ limit}.
A quite different and very interesting limit is obtained taking first the weak coupling perturbative 
expansion of $f(g,j)$ around $g=0$
\be
\label{eq:mixed}
f(g, j) = \sum_{n\ge 0} f^{(n)}(j)\,g^{2\,n}.
\ee
The functions $f^{(n)}(j)$ can be expanded around $j=0$ recovering the fully weak regime. On the other
hand, one can look for the large $j$ behavior of $f^{(n)}(j)$. This large-$j$ limit turns out to be non trivial.
Looking back at the analysis of~\cite{Belitsky:2006en}, we see that for large $M$ and fixed $j$ we simply have 
\be
\xi = \frac{1}{j}.
\ee
The result \refeq{BMNone} can be nicely rewritten in a uniform way as 
\be
f^{(1)}(j \ll 1) = 8,\qquad 
f^{(1)}(j \gg 1) = \frac{8}{j}.
\ee
These simple relations immediately suggest that the large-$j$ limit of the FRS equation is closely connected to the 
string theory calculations described first in ~\cite{Frolov:2002av} and later expanded in ~\cite{Frolov:2006qe}. In particular, the string 
perturbative 
calculations admit a BMN-like expansion which is captured by the FRS equation in the large-$j$ limit. 
The comparison can be done at {\em arbitrary coupling}, thus going beyond \refeq{mixed}. This is important 
if one is interested in detecting universal dressing effects.
In this paper, we exploit the fact that the BMN-like expansion contains some terms which are protected and can 
be computed in the weakly coupled gauge theory. This means that they can be matched by studying the large-$j$ expansion 
of the one and two-loop expansion of the FRS equation, a remarkable simplification.
Thus, we analyze the large-$j$ limit by analytical and numerical methods and provide strong support for some of the 
predictions following from the computations in~\cite{Frolov:2006qe}.

\medskip
The plan of the work is the following. In Sec.~(\ref{sec:fast}) we briefly recall a few basic facts about the so-called 
{\em fast long string} limit of the folded string solution. In Sec.~(\ref{sec:FRS}) we give a self-contained summary of the FRS
equation. In Sec.~(\ref{sec:two-loops}) we present the explicit two-loop hole density equation in Bethe roots space.
In Sec.~(\ref{sec:nnlo-one}) and Sec.~(\ref{sec:nlo-two}) we describe the analytical re-derivation of the one-loop large-$j$ limit 
at next-to-next-to-leading order and a few considerations about the similar analysis at two loops. Finally, in Sec.~(\ref{sec:numerical})
we show our numerical results supporting at two loops the predictions of~\cite{Frolov:2006qe}.

\newpage
\section{The large-$j$ limit and the fast long string limit}
\label{sec:fast}

In~\cite{Frolov:2002av,Frolov:2006qe}, S. Frolov and A. A. Tseytlin compute the semiclassical expansion 
around the rotating folded string configuration extending the analysis of~\cite{Gubser:2002tv,deVega:1996mv} and including the string 
center of mass motion along a big circle of $S^5$. Their solution depends on the Lorentz spin $M$ and $SO(6)$ 
spin $L$ to be identified with the quantum numbers of the $\mathfrak{sl}(2)$ twist operators.
The large $\lambda$ expansion of the energy takes the usual form 
\be
\label{eq:semiclassical}
E = \underbrace{\sqrt\lambda\,{\cal E}_0\left(\frac{M}{\sqrt\lambda}, \frac{L}{\sqrt\lambda}\right)}_{E_0} + 
E_1\left(\frac{M}{\sqrt\lambda}, \frac{L}{\sqrt\lambda}\right) + {\cal O}\left(\frac{1}{\sqrt\lambda}\right).
\ee
For an alternative derivation of the one-loop contribution see also~\cite{Casteill:2007ct}.
The expansion \refeq{semiclassical} can be considered in the {\em long string limit} which is
\be
\mbox{\em long string}\qquad \mbox{both}\ 1\ll \frac{M}{\sqrt\lambda} \ \mbox{and}\   \frac{L}{\sqrt\lambda} \ll \frac{M}{\sqrt\lambda}.
\ee
This limit can be further refined in the two sub-cases defined by the additional conditions
\ba
\mbox{\em slow long string} && \frac{L}{\sqrt\lambda} \ll \log\frac{M}{\sqrt\lambda}, \\
\mbox{\em fast long string} && \log\frac{M}{\sqrt\lambda} \ll \frac{L}{\sqrt\lambda}.
\ea
An interpolating regime between these two cases is obtained by fixing the parameter
\be
x = \frac{\sqrt\lambda}{\pi\,L}\,\log\frac{M}{L}.
\ee
The slow long string limit is reproduced for $x\gg 1$, the fast string limit for $x\ll 1$.
The expansion of the energy in this second limit reads~\cite{Frolov:2006qe} (see~\cite{Gromov:2008en}
for a recent analysis of $E_2$ mainly at $x\ll 1$)
\ba
E_0(x\ll 1) &=& M+L+\frac{\lambda}{2\,\pi^2\,L}\,\log^2\frac{M}{L}
-\frac{\lambda^2}{8\,\pi^4\,L^3}\,\log^4\frac{M}{L}
+\frac{\lambda^3}{16\,\pi^6\,L^5}\,\log^6\frac{M}{L}+\cdots \nonumber\\
E_1(x\ll 1) &=&  -\frac{4\,\lambda}{3\,\pi^3\,L^2}\,\log^3\frac{M}{L}
+\frac{4\,\lambda^2}{5\,\pi^5\,L^4}\,\log^5\frac{M}{L}
+\frac{\lambda^{5/2}}{3\,\pi^6\,L^5}\,\log^6\frac{M}{L}+\cdots.
\ea
The above result is a string calculation based on the large $\lambda$ assumption. However, as discussed in ~\cite{Frolov:2006qe,Beisert:2005cw,Beisert:2006ez}, 
a general form of $E$ interpolating between weak and strong coupling is expected to take the form 
\be
E = M+L+L\,\sum_{n\ge 1}\,\sum_{m\ge 0} c_{n m}(\lambda)\,\lambda^n\,\left(\frac{1}{L}\,\log\frac{M}{L}\right)^{2\,n+m}.
\ee
The coefficients $c_{n,m}(\lambda)$ have a regular expansion around $\lambda=0$ and a strong coupling expansion 
in inverse powers of $\sqrt\lambda$.
Quite remarkably, some of them are {\em protected} and are thus genuine constants independent on $\lambda$. 
This follows from the comparison between string theory and gauge theory 
of the 1-loop and 2-loop leading and subleading 
corrections~\cite{Beisert:2005mq,Hernandez:2005nf,SchaferNameki:2005tn,Gromov:2005gp,Minahan:2005mx,Minahan:2005qj}
to the thermodynamical limit of similar circular string solutions.
In particular, this is true for the coefficients
\be
c_{10},\ c_{11},\ c_{12},\ c_{20}, \ \mbox{and}\ c_{21},
\ee
and we can write the very explicit expansion (Eq.~(1.15) of ~\cite{Frolov:2006qe})
\ba
\label{eq:interpolating}
\lefteqn{E = M+L\,\big[1+} && \\
&& \frac{\lambda}{L^2}\,\log^2\frac{M}{L}\big(c_{10}+\frac{c_{11}}{L}\,\log\frac{M}{L}+
\frac{c_{12}}{L^2}\,\log^2\frac{M}{L}+\cdots\big)+ \nonumber \\
&& +\frac{\lambda^2}{L^4}\,\log^4\frac{M}{L}\big(c_{20}+\frac{c_{21}}{L}\,\log\frac{M}{L}+
\frac{c_{22}(\lambda)}{L^2}\,\log^2\frac{M}{L}+\cdots\big)+ \nonumber \\
&& +\frac{\lambda^3}{L^6}\,\log^6\frac{M}{L}\big(c_{30}(\lambda)+\frac{c_{31}(\lambda)}{L}\,\log\frac{M}{L}+
\frac{c_{32}(\lambda)}{L^2}\,\log^2\frac{M}{L}+\cdots\big)\big]+\cdots , \nonumber 
\ea
A comparison with the one-loop string energy gives ($c_{1,2}$ would require $E_2$ in the fast long string limit)
\be
\begin{array}{ll}
c_{10} = & \displaystyle +\frac{1}{2\,\pi^2}, \\ \\
c_{11} = & \displaystyle -\frac{4}{3\,\pi^3},
\end{array}
\qquad
\begin{array}{ll}
c_{20} = & \displaystyle -\frac{1}{8\,\pi^4},\\ \\
c_{21} = & \displaystyle +\frac{4}{5\,\pi^5}.
\end{array}
\ee
The value $c_{30}(0)$ is discussed in~\cite{Frolov:2006qe} and is obtained by consistency of the 
string result with the universal dressing phase~\cite{Frolov:2006qe}. It should be $c_{30}(0) = \frac{1}{8\,\pi^6}$.

Now, the crucial point is that we can take the generalized limit \refeq{limit}
in the above interpolating expansion \refeq{interpolating}. Doing so and assuming the above values for the
protected coefficients we find the following prediction for the large-$j$ behavior of the FRS generalized scaling function
\ba
\label{eq:resone}
f^{(1)}(j) &=&\frac{8}{j}-\frac{64}{3\,\pi}\,\frac{1}{j^2}+\cdots, \\
\label{eq:restwo}
f^{(2)}(j) &=& -\frac{32}{j^3}+\frac{1024}{5\,\pi}\,\frac{1}{j^4}+\cdots, \\
\label{eq:resthree}
f^{(3)}(j) &=& \frac{512}{j^5}+\cdots .
\ea
The two terms in \refeq{resone} have actually been already obtained in~\cite{Belitsky:2006en} by working out the finite size corrections to the 
semiclassical expansion of the $\mathfrak{sl}(2)$ invariant one-loop spin chain. The other two expansions are a higher loop test 
of the AdS/CFT correspondence and have not yet been computed in the gauge theory. We now discuss the confirmation of Eqs.~(\ref{eq:resone}, \ref{eq:restwo}) 
in the context of the large-$j$ FRS integral equation.

\section{The FRS equation in brief}
\label{sec:FRS}

\subsection{Setup}

We consider $\mathfrak{sl}(2)$ scaling operators of the form 
\be
\label{eq:scaling}
{\cal O} = \mbox{Tr}\,(D^M\,Z^L)+\cdots~,
\ee
where $D$ is a specific component of the covariant derivative and $Z$ a scalar field of ${\cal N}=4$ SYM~\cite{Beisert:2003jj}.
The omitted terms are analogous operators with the same number of derivatives and scalar fields. They are required to form 
an eigenstate of the dilatation operator. As usual, $L$ is also identified with the twist, {\em i.e.} the classical dimension minus
the Lorentz spin, here equal to $M$.

The anomalous dimensions of scaling operators of the form \refeq{scaling} can be organized in irreducible
multiplets of the $\mathfrak{sl}(2)$ algebra and the top states fill a band, see for instance~\cite{Belitsky:2003ys,Beccaria:2008pp}.
We can split the scaling dimension $\Delta(g)$ separating out the classical dimension and define the anomalous dimension
$\gamma(g)$ as 
\be
\Delta(g) = L+M+\gamma(g).
\ee
In terms of the energy $E(g)$ of the 
$\mathfrak{sl}(2)\subset\mathfrak{psu}(2,2|4)$ long-range integrable spin chain~\cite{Beisert:2005fw}, we have 
\be
\gamma(g) = 2\,g^2\,E(g).
\ee
The quantity $E(g)$ is the energy level of an integrable system. Therefore, it is computed by solving Bethe Ansatz
equations with suitable mode numbers identifying the relevant state in the above band.

In the FRS limit \refeq{limit}, the Bethe roots $u$ of the {\em minimal} state in the band are described by a continuous distribution with 
density $\rho_{\rm m}(u)$ supported in the region $|u|\ge c(g,j)$, where $c$ is some function of $g$ and $j$ that
we shall call {\em gap} in the following.
The label $m$ in $\rho_{\rm m}(u)$
stands for {\em magnons}, which is the standard name for the excitations of the integrable chain. Actually, the relevant 
quantity in the FRS limit is a specific contribution to $\rho_{\rm m}(u)$ called $\sigma(u)$ in \cite{Freyhult:2007pz}
and representing a fluctuation component of the magnon density.

Remarkably, there is a dual description in terms of the complementary Bethe roots called usually {\em holes}.
In the FRS limit, the holes are also described by a continuous distribution with
density $\rho_{\rm h}(u)\equiv \rho(u)$ supported in the complementary region $|u|\le c(g,j)$.
The two dual descriptions are fully equivalent and can be connected by the simple relation
\be
j\,\rho(u)=\frac{2}{\pi}-8\,\sigma(u).
\ee
The FRS equation is an all-order integral equation for the fluctuation density of magnons $\sigma(u)$. 
It can be turned into an integral equation for the hole density $\rho(u)$. We shall show that this latter equation
is better suited for the large-$j$ expansion, at least at the two loop level at which we work.

\subsection{The all-loop FRS equation}

We need a few definitions in order to write down the FRS equation. They are fully discussed in~\cite{Freyhult:2007pz}
and we summarize them here for completeness.

First, we define the BES kernel 
\be
K(t,t') = K_0(t,t')+K_1(t,t')+K_d(t,t'),
\ee
where ($J_n(t)$ is the $n$-th Bessel function)
\ba
K_0(t,t') &=& \frac{t\,J_1(t)\,J_0(t')-t'\,J_0(t)\,J_1(t')}{t^2-{t'}^2}, \\
K_1(t,t') &=& \frac{t'\,J_1(t)\,J_0(t')-t\,J_0(t)\,J_1(t')}{t^2-{t'}^2}, \\
K_d(t,t') &=& 8\,g^2\,\int_0^\infty dt''\,K_1(t, 2\,g\,t'')\,\frac{t''}{e^{t''}-1}\,K_0(2\,g\,t'', t').
\ea
In this paper, we shall not need the dressing kernel $K_d$. Then, we define also the {\em hole} kernel
\ba
K_h(t, t'; c) &=& \frac{e^\frac{t'-t}{2}}{4\,\pi\,t}\,\int_{-c}^c du\,\cos(t\,u)\,\cos(t'\,u) = \\
&=& \frac{e^\frac{t'-t}{2}}{2\,\pi\,t}\,\frac{t\,\cos(c\,t')\,\sin(c\,t)-t'\,\cos(c\,t)\,\sin(c\,t')}{t^2-{t'}^2}.\nonumber
\ea
Finally, we define the full kernel $\cal K$ as 
\ba
{\cal K}(t, t') &=& g^2\,K(2\,g\,t, 2\,g\,t')+K_h(t, t'; c)-\frac{J_0(2\,g\,t)}{t}\,\frac{\sin\,c\,t'}{2\,\pi\,t'}\,e^{\frac{t'}{2}} \\
&& -4\,g^2\,\int_0^\infty dt''\,t''\,K(2\,g\,t, 2\,g\,t'')\,K_h(t'', t'; c).\nonumber
\ea
After these preliminary definitions we are ready to write the FRS equation which holds for the 
Fourier transform of the (even) magnon fluctuation density $\sigma(u)$
\be
\hat{\sigma}(t) = e^{-\frac{t}{2}}\int_\mathbb{R} du\,e^{-i\,t\,u}\,\sigma(u) = 2\, e^{-\frac{t}{2}}\int_0^\infty du\,\cos(t\,u)\,\sigma(u).
\ee
The all-loop FRS equation reads
\be
\label{eq:FRS}
\hat{\sigma}(t) = \frac{t}{e^t-1}\left({\cal K}(t, 0)-4\int_0^\infty dt'\,{\cal K}(t, t')\,\hat{\sigma}(t')\right).
\ee
The $j$ parameter is related to the gap parameter $c$ by the relation
\be
\label{eq:j}
j = \frac{4\,c}{\pi}-\frac{16}{\pi}\int_0^\infty dt\,\frac{\sin\,c\,t}{t}\,e^{\frac{t}{2}}\,\hat{\sigma}(t).
\ee

The generalized scaling function $f(g,j)$ has a rather complicated all-loop expression in terms of the
solution to the FRS equation
\ba
\lefteqn{f(g,j) = 8\,g^2\,\left[1-8\int_0^\infty dt\frac{J_1(2gt)}{2gt}\,t\,K_h(t,0; c)\right. } && \\
&& \left. -8\int_0^\infty dt\,\frac{J_1(2gt)}{2gt}\left(
\sigma(t)-4\,t\,\int_0^\infty dt' K_h(t,t'; c)\,\hat\sigma(t')\right)\right]. \nonumber
\ea
As shown in~\cite{Freyhult:2007pz}, this can also be written more simply
as 
\be
f(g, j) = j + 16\,\hat{\sigma}(0).
\ee

\section{The two-loop hole density equation in $u$-space}
\label{sec:two-loops}

The FRS equation can be rewritten in $u$-space by Fourier analyzing \refeq{FRS}. We did the analysis up to the two loop level.
After some manipulations we arrive at the following result where we use the notation of Appendix~(\ref{app:psi}) for the $G_a$ functions
\ba
\label{eq:FRS2}
\rho(u) &=& \frac{2}{\pi\,j} -\frac{1}{2\,\pi}\,G_{1/2}(u) + \int_{-c}^c\frac{dv}{2\,\pi}\,G_0(u-v)\,\rho(v) + \\
&& + g^2\,\left[
-\frac{1}{2\,\pi}\,G''_{1/2}(u)-\frac{\pi}{4\,j\,\cosh^2(\pi\,u)}\,\gamma_1[\rho^{(0)}]\right] + \oh{g^4}.\nonumber
\ea
In this equation, $\rho^{(0)}$ is the one-loop term in the weak coupling expansion of the hole density
\be
\label{eq:expansion}
\rho(u) = \rho^{(0)}(u) + g^2\, \rho^{(1)}(u) + \oh{g^4},
\ee
and the functional $\gamma_1[\rho]$ is defined as 
\be
\gamma_1[\rho] = 8+2\,j\,\int_{-c}^c \left[G_{1/2}(u)+2\,\gamma_{\rm E}\right]\,\rho(u)\,du.
\ee
The generalized scaling function turns out to have the following explicit expression
\ba
\label{eq:gamma2}
\lefteqn{f(g, j) = g^2\,\gamma_1[\rho]+}&& \\
&& + g^4\left[8\,j\,\left(-\zeta_3-\frac{\pi^2}{24\,j}\,\gamma_1[\rho^{(0)}]\right) -j\,\int_{-c}^c \left[-G''_{1/2}(u)+4\,\zeta_3\right]\,\rho(u)\,du\right] + \oh{g^6}. \nonumber
\ea
The one-loop terms in Eqs.~(\ref{eq:FRS2}, \ref{eq:gamma2}) are of course identical to those already written in~\cite{Freyhult:2007pz}.
As a check of the two-loops terms, one can compute the small $j$ expansion of \refeq{FRS2} and reproduces perfectly the results of~\cite{Freyhult:2007pz} for the generalized scaling function 
at two loops. Also, the two-loop expression of the gap in this regime is 
\ba
c(g,j) &=& j\,\left(\frac{\pi}{4}+g^2\,\frac{\pi^3}{4}\right) + \\
&& + j^2\,\left(-\frac{\pi}{4}\,\log\,2+\frac{g^2}{4}\left(-3\,\pi^3\,\log\,2+7\,\pi\,\zeta_3\right)\right) + \nonumber\\
&& + j^3\,\left(\frac{\pi}{4}\,\log^2\,2+\frac{g^2}{192}\left(-\pi^7+240\,\pi^3\,\log^2\,2-672\,\pi\,\log\,2\,\zeta_3\right)\right) + \nonumber\\
&& + \oh{j^4}.\nonumber
\ea

\section{NNLO large-$j$ expansion of the one-loop FRS equation}
\label{sec:nnlo-one}

We now work out the next-to-next-to-leading (NNLO) large-$j$ expansion of the one-loop hole density equation and generalized scaling function. This is 
feasible and gives the value of the leading term in \refeq{resone}. This is a confirmation of the
calculation described in~\cite{Belitsky:2006en} obtained independently in the large-$j$ FRS context. 
Also, we find additional information on the density profile as well as on the dependence of the gap on $j\gg 1$.

Using the notation of Appendix~(\ref{app:psi}), the one-loop hole density $\rho(u)$ satisfies the equation 
\be
\rho(u) = \frac{2}{\pi\,j}-\frac{1}{2\,\pi}\,G_{1/2}(u) + \frac{1}{2\,\pi}\, \int_{-c}^c dv\,G_0(u-v)\,\rho(v),
\ee
with the normalization condition relating $c$ and $j$
\be
\int_{-c}^c \rho(u)\,du = 1.
\ee
The one loop contribution to the generalized scaling function is simply ($\psi(z) = \frac{d}{dz}\,\log\,\Gamma(z)$)
\be
f^{(1)}(j) = 8 + 2\,j\,\int_{-c}^c du\,\left[G_{1/2}(u)-2\,\psi(1)\right]\,\rho(u).
\ee

\subsection{Leading order}

If $j\to\infty$ we expect $c\to\infty$. In this limit and using the normalization condition, we can write the density equation as
\be
\rho(u) = \frac{1}{2\,\pi}\, \int_{-\infty}^\infty dv\,\left[G_0(u-v)-G_{1/2}(u)\right]\,\rho(v).
\ee
We now recall the useful  integral representation 
\be
\int_0^\infty dt\, \frac{e^{\frac{t}{2}}-\cos(t\,y)}{e^t-1}\,\cos(t\,x) =
 \frac{1}{4}\left(G_0(x-y)+G_0(x+y)-2\,G_{1/2}(x)\right).
\ee
Since $\rho(u) = \rho(-u)$, we obtain 
\be
\rho(u) = \frac{1}{\pi} \int_{-\infty}^\infty dv\,\int_0^\infty dt\,\frac{e^{\frac{t}{2}}-\cos(t\,v)}{e^t-1}\,\cos(t\,u)\,\rho(v)
\ee
Introducing the Fourier transform
\ba
\widetilde\rho(t) &=& \int_{-\infty}^\infty du\,e^{i\,t\,u}\,\rho(u) =  \int_{-\infty}^\infty du\, \cos(t\,u)\,\rho(u), \\
\rho(u) &=& \frac{1}{2\,\pi}\,\int_{-\infty}^\infty dt\,e^{-i\,t\,u}\,\widetilde\rho(t) =  \frac{1}{\pi}\,\int_0^\infty dt\, \cos(t\,u)\,\widetilde\rho(t),
\ea
we find
\be
\rho(u) = \frac{1}{\pi} \int_0^\infty dt\,\frac{e^{\frac{t}{2}}-\widetilde\rho(t)}{e^t-1}\,\cos(t\,u).
\ee
This means that 
\be
\widetilde\rho(t) = \frac{e^{\frac{t}{2}}-\widetilde\rho(t)}{e^t-1},\qquad\la\qquad \widetilde\rho(t) = e^{-\frac{|t|}{2}},
\ee
where we do not restrict $t$ to be positive. Hence, at leading order, 
\be
\rho(t) = \frac{1}{2\,\pi}\,\frac{1}{u^2 + \frac{1}{4}}.
\ee
The contribution to the scaling function is easily evaluated. We start from 
\be
G_{1/2}(u)-2\,\psi(1) = 2\,\int_0^\infty dt\,\frac{e^{-t}-e^{-\frac{t}{2}}\,\cos(t\,u)}{1-e^{-t}},
\ee
and compute
\be
\int_{-\infty}^\infty du\,\int_0^\infty dt\,\frac{e^{-t}-e^{-\frac{t}{2}}\,\cos(t\,u)}{1-e^{-t}}\,\frac{1}{2\,\pi}\,\frac{1}{u^2 + \frac{1}{4}} = 
\int_0^\infty dt\,\frac{e^{-t}-e^{-\frac{t}{2}}\,e^{-\frac{t}{2}}}{1-e^{-t}} = 0.
\ee
This shows that 
\ba
f^{(1)}(j) &=& 0\cdot j + {\cal O}(1), \\
c &=& {\cal O}(j).
\ea
and the dominant term linear in $j$ cancels. The constant $8$ contribution to $f^{(1)}(j)$ is included in the $\oh{1}$ terms.

\subsection{Next-to-leading order}

The expansion at large $j$ is not trivial and is better performed in $u$-space. Let us write
\be
\rho(u) = \rho_0(u) + \delta\rho(u),\qquad \rho_0(u) = \frac{1}{2\,\pi}\,\frac{1}{u^2 + \frac{1}{4}}.
\ee
The density equation can be written 
\be
\delta\rho(u) = \frac{2}{\pi\,j}-\int_{|v|>c}\frac{dv}{2\,\pi}\, G_0(u-v)\,\rho_0(v) + \int_{-c}^c \frac{dv}{2\,\pi}\, G_0(u-v)\,\delta\rho(v).
\ee
Rescaling $u=c\,x$, $v=c\,y$, the first integral can be uniformly expanded as
\be
\int_{|v|>c}\frac{dv}{2\,\pi}\, G_0(u-v)\,\rho_0(v) = \frac{\log\,c}{c\,\pi^2} + \frac{1}{c}\,\Phi(x) + {\cal O}\left(\frac{1}{c^3}\right),
\ee
where
\be
\Phi(x) = \frac{(1+x)\,\log(1+x)-(1-x)\,\log(1-x)}{2\,\pi^2\,x}.
\ee
This suggest to set 
\be
\delta\rho(u) = \frac{1}{c^2}\,\rho_1\left(\frac{u}{c}\right)+\cdots~.
\ee
The density equation is then
\be
\frac{2\,c}{\pi\,j}-\frac{\log\,c}{\pi^2}-\Phi\left(\frac{u}{c}\right) + \frac{1}{c}\,\int_{-c}^c \frac{dv}{2\,\pi}\,G_0(u-v)\,\rho_1\left(\frac{u}{c}\right) = 0.
\ee
In terms of $x, y$ it is 
\be
\frac{2\,c}{\pi\,j}-\frac{\log\,c}{\pi^2}-\Phi(x) + \int_{-1}^1 \frac{dy}{2\,\pi}\,G_0(c\,(x-y))\,\rho_1(y) = 0.
\ee
Using
\ba
G_0(c\,x) &=& 2\,\log\,c + 2\,\log|x| + \cdots, \\
c &=& \alpha\,j + \cdots,
\ea
with an undetermined constant $\alpha$, we arrive at the singular problem
\ba
\label{eq:singularproblem}
\frac{1}{\pi}\int_{-1}^1 dy\,\log|x-y|\,\rho_1(y) &=& \Phi(x)-\frac{2\,\alpha}{\pi}, \\
\label{eq:normsing1}
\int_{-1}^1 dy\,\rho_1(y) &=& \frac{1}{\pi}.
\ea
Notice that the normalization of $\rho_1$ is also consistently predicted by 
\be
1 = \int_{-c}^c du\,\left[\rho_0(u) + \frac{1}{c^2}\, \rho_1\left(\frac{u}{c}\right)\right] + {\cal O}\left(\frac{1}{c^2}\right),
\ee
which, evaluating the elementary integral involving $\rho_0$, reads
\be
1 = 1 + \frac{1}{c}\left[-\frac{1}{\pi} +  \int_{-1}^1 dx\,\rho_1(x)\right] +  {\cal O}\left(\frac{1}{c^2}\right),
\ee
and leads to \refeq{normsing1}.
The constant $\alpha$ can be determined dividing by $\sqrt{1-x^2}$ and integrating. The key result 
\be
\int_{-1}^1 dx\,\frac{\log|x-y|}{\sqrt{1-x^2}} = -\pi\,\log\,2,
\ee
together with the elementary integral
\be
\int_{-1}^1\frac{dx}{\sqrt{1-x^2}} = \pi, 
\ee
as well as 
\be
\int_{-1}^1 dx\,\frac{\Phi(x)}{\sqrt{1-x^2}} = \frac{1}{2}-\frac{\log\,2}{\pi},
\ee
leads to 
\be
-\frac{\log\,2}{\pi} = \frac{1}{2}-\frac{\log\,2}{\pi}-2\,\alpha,\qquad \la\qquad \alpha = \frac{1}{4}.
\ee
The solution of the integral equation with logarithmic kernel is standard. By taking a derivative it is reduced to a finite Hilbert transform 
problem and the solution is~\cite{SingBook}
\be
\rho_1(x) = \frac{1}{\pi^2}\,\frac{1}{\sqrt{1-x^2}} \, \left[1 + \pint_{-1}^1 dy\,\sqrt{1-y^2}\,\frac{\pi\,\Phi'(y)}{y-x}  \right]
\ee
Evaluating the principal integral we get
\be
\label{eq:rho1}
\rho_1(x) = \frac{1}{2\,\pi}\,\frac{1}{\sqrt{1-x^2}}\,\frac{1}{1+\sqrt{1-x^2}}.
\ee

The evaluation of the generalized scaling function can be done as follows. We want to compute
\ba
\lefteqn{\int_{-c}^c du\,\left[G_{1/2}(u)+2\,\gamma_E\right]\,\left(\rho_0(u) + \frac{1}{c^2}\,\rho_1(u/c) + \cdots\right) = } && \\
&& \int_{-c}^c du\,\left[G_{1/2}(u)+2\,\gamma_E\right]\,\rho_0(u) + \frac{1}{c}\int_{-1}^1 dx\,\left[G_{1/2}(c\,u)+2\,\gamma_E\right]\,\rho_1(x) + \cdots \nonumber
\ea
The first integral reads
\be
F(c) = \int_{-c}^c du\,\frac{\psi(\frac{1}{2}+i\,u)+\psi(\frac{1}{2}-i\,u)+2\,\gamma_E}{2\,\pi\,(u^2+\frac{1}{4})}.
\ee
We know that $F(\infty) = 0$. Also, 
\be
F'(c) = \frac{\psi(\frac{1}{2}+i\,c)+\psi(\frac{1}{2}-i\,c)+2\,\gamma_E}{\pi\,(c^2+\frac{1}{4})} = \frac{2}{\pi\,c^2}(\log\,c+\gamma_E) + {\cal O}\left(\frac{1}{c^4}\right).
\ee
Hence, 
\be
F(c) = -\frac{2}{\pi\,c}(1+\gamma_E+\log\,c) +  {\cal O}\left(\frac{1}{c^3}\right).
\ee
The second integral is for large $c$
\ba
\lefteqn{
\int_{-1}^1 dx\,\left[G_{1/2}(c\,u)+2\,\gamma_E\right]\,\rho_1(x) = }&& \\
&& 2\,(\gamma_E+\log\,c)\,\int_{-1}^1 dx\,\rho_1(x) + 2\,\int_{-1}^1 dx\,\log|x|\,\rho_1(x) = \\
&& \frac{2\,(\gamma_E+\log\,c)}{\pi}+2\,\pi\left(\Phi(0)-\frac{1}{2\,\pi}\right) = \frac{2\,(\gamma_E+\log\,c)}{\pi}+\frac{2}{\pi}-1.
\ea
Combining, we find
\be
\int_{-c}^c du\,\left[G_{1/2}(u)+2\,\gamma_E\right]\,\left(\rho_0(u) + \frac{1}{c^2}\,\rho_1(u/c) + \cdots\right) = -\frac{1}{c}
+{\cal O}\left(\frac{1}{c^3}\right).
\ee
We conclude that
\ba
f^{(1)}(j) &=& 8 + 0\cdot j + \frac{2}{\alpha} + {\cal O}\left(\frac{1}{c}\right) = 0\cdot j + 0 + {\cal O}\left(\frac{1}{j}\right), \\
c &=& \frac{1}{4}\,j + \mbox{subleading}.
\ea

\subsection{Next-to-next-to-leading order}

Expanding further the density equation with the position 
\be
\delta\rho(u) = \frac{1}{c^2}\,\rho_1\left(\frac{u}{c}\right) + \frac{1}{c^3}\,\rho_2\left(\frac{u}{c}\right) + \cdots,
\ee
and using the results in Appendix, we find 
\ba
\lefteqn{\frac{1}{c^2}\,\rho_1(x) + {\cal O}\left(\frac{1}{c^3}\right) = \frac{2}{\pi\,j}-\frac{\log\,c}{\pi^2\,c}-\frac{1}{c}\,\Phi(x) + {\cal O}\left(\frac{1}{c^3}\right) + } && \\
&& +c\,\int_{-1}^1\frac{dy}{\pi}\left(\log\,c+\log|x-y|+\frac{\pi}{2\,c}\,\delta(x-y) + \cdots\right)\left(
\frac{1}{c^2}\,\rho_1(y) + \frac{1}{c^3}\,\rho_2(y) + \cdots\right)\nonumber
\ea
We assume the following general expansion of the gap
\be
c = \frac{j}{4}+\beta\,\log\,j+\gamma + \cdots, 
\ee
Hence
\be
j = 4\,c-4\,\beta\,\log\,c-8\,\beta\,\log\,2-4\,\gamma + \cdots.
\ee
and
\be
\frac{2}{\pi\,j} = \frac{1}{2\,\pi\,c}+\left(\frac{\gamma}{2\,\pi}+\frac{\beta\,\log\,2}{\pi} + \frac{\beta\,\log\,c}{2\,\pi}\right)\frac{1}{c^2} + \cdots.
\ee
The normalization condition 
\be
1 = \int_{-c}^c du\,\left[\rho_0(u) + \frac{1}{c^2}\, \rho_1\left(\frac{u}{c}\right)+ \frac{1}{c^3}\, \rho_2\left(\frac{u}{c}\right)\right] + {\cal O}\left(\frac{1}{c^4}\right),
\ee
leads to 
\be
\int_{-1}^1 dx\,\rho_2(x) = 0.
\ee
The relevant terms in the density equation are thus
\be
\frac{1}{\pi}\int_{-1}^1 dx\,\log|x-y|\,\rho_2(y) = \frac{1}{2}\,\rho_1(x)-\frac{\gamma}{2\,\pi}-\frac{\beta\,\log\,2}{\pi}-\frac{\beta\,\log\,c}{2\,\pi}.
\ee
The appearance of the logarithmic term $\log\,c$ is tricky and can be understood as follows. Dividing the above equation by $\sqrt{1-x^2}$ and 
integrating between $-1$ and $1$ we find using the normalization of $\rho_2$
\be
\label{eq:normsing}
0 = \frac{1}{2}\,\int_{-1}^1 dx\,\frac{\rho_1(x)}{\sqrt{1-x^2}}-\frac{\gamma}{2}-\beta\,\log\,2-\frac{\beta\,\log\,c}{2} .
\ee
On the other hand, the $\rho_1$ function has the following behavior for $|x|\to 1$
\be
\rho_1(x)\sim \frac{1}{2\,\pi\sqrt{1-x^2}}.
\ee
Thus the integral in \refeq{normsing} is singular
\be
\frac{1}{2}\,\int_{-1}^1 dx\,\frac{\rho_1(x)}{\sqrt{1-x^2}} = \frac{1}{4\,\pi}\int_{-1}^1\frac{dx}{1-x^2} = \infty.
\ee
The most singular and universal term is evaluated by integrating over $[-1+\varepsilon, 1-\varepsilon]$ and identifying $\varepsilon\sim 1/c$.
The logarithmic singularity is not ambiguous and reads
\be
\frac{1}{2}\,\int_{-1}^1 dx\,\frac{\rho_1(x)}{\sqrt{1-x^2}} = \frac{\log\,c}{4\,\pi} + \mbox{less singular}.
\ee
This gives
\be
\label{eq:heuristic}
\beta = \frac{1}{2\,\pi}.
\ee
Apart from this, we shall not attempt to determine more precisely the function $\rho_2$ nor the constant $\gamma$ which we shall not need in the end.
Instead, we notice the important relation
\be
\frac{1}{\pi}\int_{-1}^1 dx\,\log|y|\,\rho_2(y) = \frac{1}{2}\,\rho_1(0)-\frac{\gamma}{2\,\pi}-\frac{\beta\,\log\,2}{\pi}-\frac{\beta\,\log\,c}{2\,\pi}
\ee
Computing the central value
\be
\rho_1(0) = \frac{1}{4\,\pi},
\ee
we thus obtain a crucial piece in the determination of the anomalous dimension at NNLO
\be
\int_{-1}^1 dx\,\log|y|\,\rho_2(y) = \frac{1}{8}-\frac{\gamma}{2}-\beta\,\log\,2-\frac{\beta\,\log\,c}{2}.
\ee
If we look back to the expression determining the anomalous dimension, we see that all the terms from $\rho_0$ are already at the precision 
of NNLO. Also, the integral involving $\rho_1$ is 
\be
\int_{-1}^1 dx\,\left[G_{1/2}(c\,x)+2\,\gamma_E\right]\,\rho_1(x)
\ee
Due to the results in the Appendix, the expansion of $G_{1/2}$ has no $\delta$-term and is also already at the NNLO precision. 
The missing piece is (using again the normalization of $\rho_2$)
\ba
\lefteqn{
\int_{-1}^1 dx\,\left[G_{1/2}(c\,x)+2\,\gamma_E\right]\,\rho_2(x) = }&& \\
&& 2\,(\gamma_E+\log\,c)\,\int_{-1}^1 dx\,\rho_2(x) + 2\,\int_{-1}^1 dx\,\log|x|\,\rho_2(x) = \\
&& 2\,\int_{-1}^1 dx\,\log|x|\,\rho_2(x) = \frac{1}{4}-\gamma-2\,\beta\,\log\,2-\beta\,\log\,c.
\ea
Expanding in powers of $j$ the combination 
\be
8+2\,j\,\left(-\frac{1}{c} + \frac{1}{c^2}\left(\frac{1}{4}-\gamma-2\,\beta\,\log\,2-\beta\,\log\,c\right)\right) + {\cal O}\left(\frac{1}{c^3}\right),
\ee
we find 
\ba
\label{eq:quizzy}
f^{(1)}(j) &=& 0\cdot j + 0 + \frac{8}{j} + {\cal O}\left(\frac{1}{j^2}\right), \\
c &=& \frac{1}{4}\,j + \frac{1}{2\,\pi}\,\log\,j + {\cal O}(1).
\ea
Notice also that \refeq{quizzy} is completely independent on both $\beta$ and $\gamma$ that, to be honest, deserve a full determination 
at next order.

\section{Large-$j$ expansion of the two loops FRS equation: LO and remarks}
\label{sec:nlo-two}

The leading order calculation is very simple. We start from the surviving terms for $j\to\infty$ and at this order
we can set $c=\infty$. The equation to be solved is
\be
\rho(u) = \frac{1}{2\,\pi}\, \int_{-\infty}^\infty dv\,\left[G_0(u-v)-G_{1/2}(u)-g^2\,G''_{1/2}(u)\right]\,\rho(v).
\ee
A Fourier analysis completely similar to the one discussed for the one-loop case gives the interesting result
\be
\rho(u) = \frac{2}{\pi\,(1+4\,u^2)}-\frac{16}{\pi}\,\frac{1-12\,u^2}{(1+4\,u^2)^3}\,g^2 + \oh{g^4}.
\ee
The integral of the two-loop correction vanishes and one easily proves the cancellation of $\oh{j}$ terms in
the expression of $f^{(2)}(j)$. The NLO correction is also similar to the one-loop case and one proves the 
cancellation of $\oh{j}$ terms in the gap as well as cancellation of $\oh{1}$ terms in $f^{(2)}(j)$.
We did not push the calculation further since we expect that $f^{(2)}(j) = \oh{1/j^3}$ which means that we 
need a $\mbox{N}^4\mbox{LO}$ calculation !
For this reason we shall discuss in the next section a fully numerical determination of the two 
terms in \refeq{restwo}. In perspective, this shows that beyond one-loop more sophisticated analytical tools
are necessary instead of the {\em brute-force} one-loop analysis.

\section{Numerical study of the hole density equation}
\label{sec:numerical}

We look for a numerical determination of the one and two loop densities in \refeq{expansion}. We need also 
to expand the gap
\be
c = c^{(0)} + g^2\,c^{(1)} + \oh{g^4}.
\ee

\subsection{One loop}

The numerical problem at one-loop is the solution of the non-singular integral equation
\ba
\rho^{(0)}(u) &=& \frac{2}{\pi\,j}-\frac{1}{2\,\pi}\,G_{1/2}(u)+\int_{-c^{(0)}}^{c^{(0)}}\frac{dv}{2\,\pi} \,G_0(u-v)\,\rho^{(0)}(v)\,dv, \\
1 &=& \int_{-c^{(0)}}^{c^{(0)}} \rho^{(0)}(u)\,du.
\ea
To this aim, we fix $c^{(0)}$ and discretize the $u$ space evaluating the integral by the Boole's rule. The integral equation becomes a linear
problem which can be solved very efficiently and with high accuracy. The resulting density is plugged in the area constraint and $c^{(0)}$ is
determined by bisection. All the procedure must be repeated with smaller and smaller lattice spacings until convergence is achieved.
The good convergence is shown in \reffig{convergence} at $j=10$.

As a first result, we show in \reffig{one-rho} the one loop hole density from the numerical integration of the FRS equation at the three values of the $j$
parameter $j=10, 30$ and $j=60$. The density is progressively better represented by the LO analytical expression as $j$ increases. One notices
that the tails of the density, near the boundary of the gap interval, show an interesting small rising. This is precisely captured by the NLO
solution as shown in \reffig{one-rho-corr}. In that figure, we show the numerical density after subtraction of the analytical LO contribution.
For the $j=30, 60$ we also superimpose the analytical NLO solution. It is a rather good approximation, although the finite $j$
data cannot show the divergence on the boundary $u=\pm c$. As a further check, we also show, in the $j=30$ panel,  the numerical
solution of the logarithmically singular equation \refeq{singularproblem} that determines $\rho_1$. The agreement with \refeq{rho1} is of course perfect. 

The dependence of the gap on $j$ is illustrated in \reffig{one-gap} where we subtract out the leading contribution $j/4$ in order to 
better display the subleading terms. Indeed, a non-trivial reminder can be seen which is very well fitted by the heuristic logarithmic term 
\refeq{heuristic} discussed previously. 

Finally, we show in \reffig{one-gamma} the numerical computation of the generalized scaling function at one-loop. The data are very well 
reproduced by the NLO prediction \refeq{resone}. We also show that the LO prediction is not enough to reproduce the numerics. This is a confirmation 
of the NLO contribution. By the way, one can also make a general 2 or 3-parameter fit to predict a priori the two coefficients and of course
they are matched with good precision below the 0.1 \% level.

\subsection{Two loops}

The two loop equation for $\rho^{(1)}$ and the constraint on $c^{(1)}$ are
\ba
\rho^{(1)}(u) &=& \int_{-c^{(0)}}^{c^{(0)}}\frac{dv}{2\,\pi} \,G_0(u-v)\,\rho^{(1)}(v)\,dv-\frac{1}{2\,\pi}\,G''_{1/2}(u)  
-\frac{\pi}{4\,j\,\cosh^2(\pi\,u)}\,\gamma_1[\rho^{(0)}] + \nonumber\\
&& + \frac{1}{2\,\pi}\,c^{(1)}\,\rho^{(0)}(c^{(0)})\,\left[G_0(u-c^{(0)})+G_0(u+c^{(0)})\right],
\ea
\be
2\,\rho^{(0)}(c^{(0)})\,c^{(1)} + \int_{-c^{(0)}}^{c^{(0)}} \rho^{(1)}(u)\,du = 0.
\ee
By discretization, the first equation is a linear problem where we have to insert various quantities computed in the solution of the 
one-loop problem. This must be done with a fixed $c^{(1)}$ which is then evaluated by bisection to impose the second constraint.

The two-loop contribution to the density profile is illustrated in \reffig{two-rho}. Apart from remarkably small corrections, the LO 
expression captures essentially the numerical data. The gap is shown in \reffig{two-gap} where we confirm that the two 
loop contribution starts $\oh{1/j}$. We have fitted the numerical results with a 3-parameter fit. The leading term is very accurately
$6/j$. In the figure, we show the result of a 2-parameter fit with the leading term fixed at that value.

Finally, in \reffig{two-gamma}, we show the two loop generalized scaling function. As in the one-loop case, one can predict the coefficients.
The result is in perfect agreement with \refeq{restwo}. This is best illustrated by superimposing the NLO curve which reproduces numerical data
very well. Again, one can try to see the accuracy of the LO term alone and the figure shows that it is not enough. This means that
our computation strongly suggest the validity of the expansion \refeq{restwo}.

\section{Conclusions}

In this paper, we have exploited a very simple remark, {\em i.e.} the observation that the large-$j$ limit of the FRS equation 
can be used to capture the fast spinning long string limit of $\ads$ superstring perturbation theory.
Indeed, in this limit, the string energy can be expanded in inverse powers of $j$
as 
\ba
E &=& M+L+f(\lambda, j)\,\log\,M + \oh{M^0}, \\
f(\lambda, j) &=& \sum_{n\ge 1}\,C_{n}(\lambda)\,j^{-n}.
\ea
In many respects, the coefficients $C_n(\lambda)$ are similar to the more studied scaling function and its generalizations 
which are obtained by expanding $f(g, j)$ around $j=0$. Indeed, the various $C_n(\lambda)$ are defined for all $\lambda$ by the above relation
and can be computed at any $\lambda$, weak or strong,  by taking the large-$j$ limit of the FRS equation.
As we explained, the large-$j$ limit is quite natural from the point of view of string perturbation theory where the two
regimes of slow and fast long string emerges quite symmetrically and are associated with $j\ll 1$ or $j\gg 1$.

As a first attempt to study the large-$j$ limit of the FRS equation, we have described in this paper what can be learned 
in the weakly coupled gauge theory. This regime has a non vanishing overlap with the string calculation since part of 
the $C_n(\lambda)$ coefficients is protected leading to a prediction from string theory valid also at small $\lambda$. 

At one-loop in the gauge theory, we can match the result \refeq{resone}. This is not a new check since \refeq{resone} has already been obtained
by working out the finite size corrections to the integrable $XXX_{-1/2}$ spin chain~\cite{Belitsky:2006en}. Nevertheless, this is an important check of the 
approach and several interesting new details are uncovered. In particular, we have obtained various results concerning the large-$j$ Bethe roots
density and gap dependence.

At two-loops in the gauge theory, we can match \refeq{restwo}. This is an interesting check first proposed in~\cite{Frolov:2006qe} and never
verified. We did not work it out in a fully analytical way, but have 
shown that a numerical approach is feasible and strongly supports a perfect agreement. In principle, an analogous study could be carried over
to test the three loop result \refeq{resthree}.

Clearly, the most interesting development of our analysis is to compute the strong coupling expansion of $C_n(\lambda)$ from the FRS equation.
It would be very interesting to investigate whether the effective techniques developed in~\cite{Basso:2007wd,Basso:2008tx} for $j\ll 1$
can also  be applied to the large-$j$ case.

\acknowledgments

We thank A.~A.~Tseytlin for kind interest and comments. We thank G. F. De Angelis and V.~Forini for useful comments.
We thank the Department of Physics of the Humboldt-Universit\"at zu Berlin for kind hospitality.  

\appendix
\section{Technical results about the combinations $\psi(a+i\,x)+\psi(a-i\,x)$}
\label{app:psi}

Let us consider the function ($\psi(z) = \frac{d}{dz}\,\log\,\Gamma(z)$)
\be
\label{eq:G}
G_a(x) = \psi(a+i\,x)+\psi(a-i\,x),\qquad a,x\in\mathbb{R},\ \mbox{and}\  a>0.
\ee
The function $G_a(x)$ is real. Due to a special reflection property of the $\psi(z)$ function, 
one has the remarkable identity
\be
G_1(x) = \psi(i\,x)+\psi(-i\,x).
\ee
Hence, it will be convenient to extend the definition \refeq{G} to the case $a=0$ 
by understanding
\be
G_0(x)\equiv G_1(x).
\ee
From the integral representation
\be
\psi(z)+\gamma_E = \int_0^\infty ds\,\frac{e^{-s}-e^{-z\,s}}{1-e^{-s}},\qquad \mbox{Re}(z)>0,
\ee
we can obtain the remarkable definite integral
\be
G_a(x)-2\,\log|x| = -2\int_0^\infty\left(\frac{e^{-a\,s}}{1-e^{-s}}-\frac{1}{s}\right)\,\cos(s\,x).
\ee
This means that the following Fourier transform holds
\be
{\cal F}\big\{G_a(x)-2\,\log|x|\big\} = -2\,\pi\,\left(\frac{e^{-a|t|}}{1-e^{-|t|}}-\frac{1}{|t|}\right).
\ee
Also, using the above reflection identity, we have for $a=0$
\be
{\cal F}\big\{G_0(x)-2\,\log|x|\big\} = -2\,\pi\,\left(\frac{1}{e^{|t|}-1}-\frac{1}{|t|}\right).
\ee
If we now want to compute the asymptotic expansion for $c\to\infty$ of the integral
\be
I = \int_{-1}^1 dx\,G_a(c\,x)\,\rho(x),
\ee
we simply add and subtract a logarithm and obtain 
\be
I = 2\,\log\,c\,\int_{-1}^1 dx\,\rho(x) + 2\,\int_{-1}^1 dx\,\log\,|x|\,\rho(x) + \int_{-1}^1 dx\,(G_a(c\,x)-\log\,(c|x|))\,\rho(x).
\ee
The {\em first} term in the asymptotic expansion of the last integral is obtained by writing it as the integral of the 
Fourier transforms of $G_a(c\,x)-\log\,(c|x|)$ and $\rho$ and expanding the above results. 
We can compactly write the result as the distributional identity
\ba
G_0(c\,x) &=& 2\,\log\,c + 2\,\log|x|+\frac{\pi}{c}\,\delta(x) + {\cal O}\left(\frac{1}{c^2}\right), \\
G_\frac{1}{2}(c\,x) &=& 2\,\log\,c + 2\,\log|x|+ {\cal O}\left(\frac{1}{c^2}\right).
\ea
The $\delta$-term in the $a=0$ case is quite important for the discussion of the main text.

\newpage
\FIGURE{
\vspace*{0.6cm}
\epsfig{file=Figures/convergence.eps,height=12cm}
\vspace*{0.5cm}
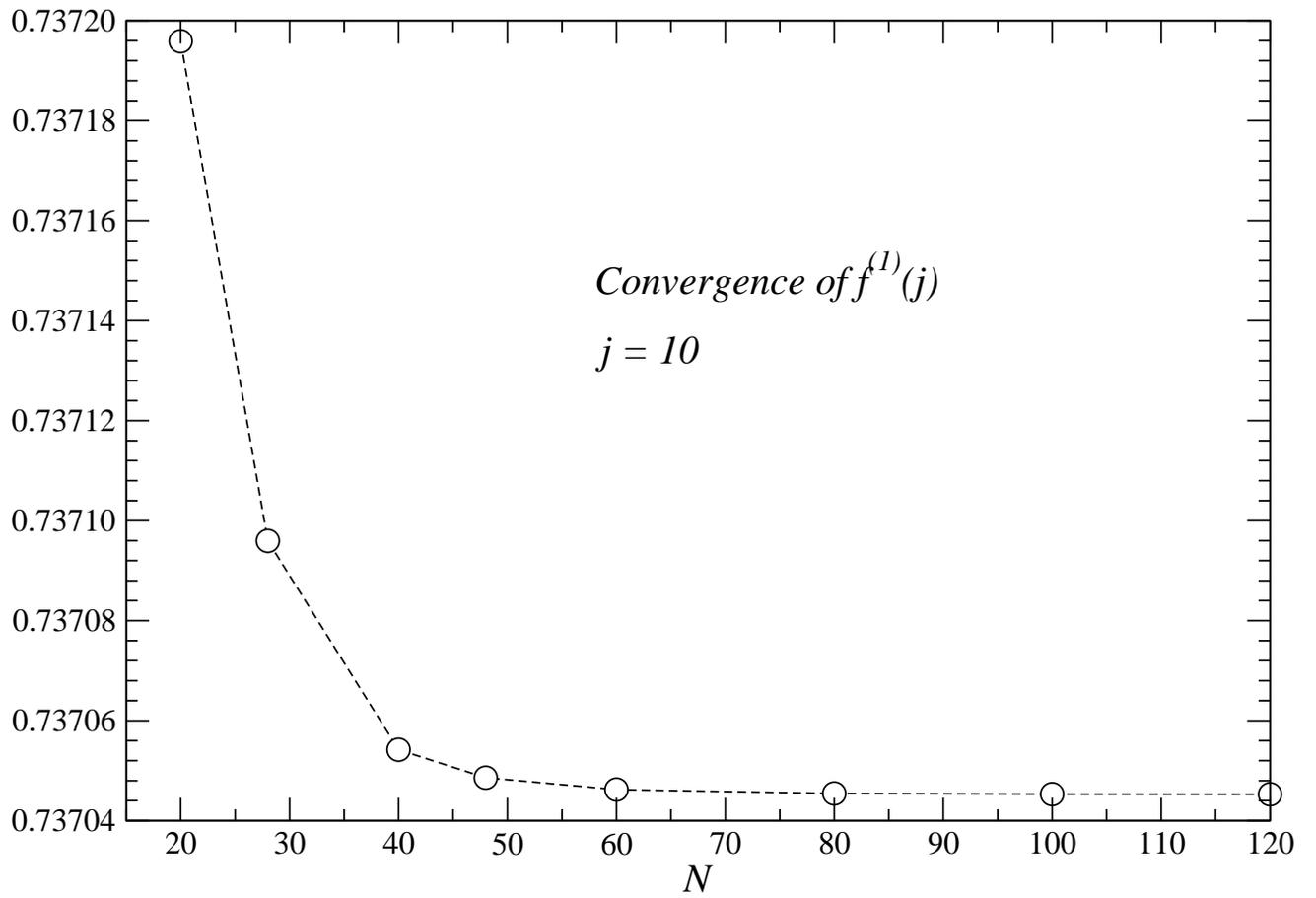
\caption{Convergence of the one-loop contribution $f^{(1)}(j)$ at $j=10$ as the number of discretized points $N$ in $u$-space is increased.}
\label{fig:convergence}
}

\newpage
\FIGURE{
\vspace*{0.6cm}
\epsfig{file=Figures/oneloop.rho.eps,height=12cm}
\vspace*{0.5cm}
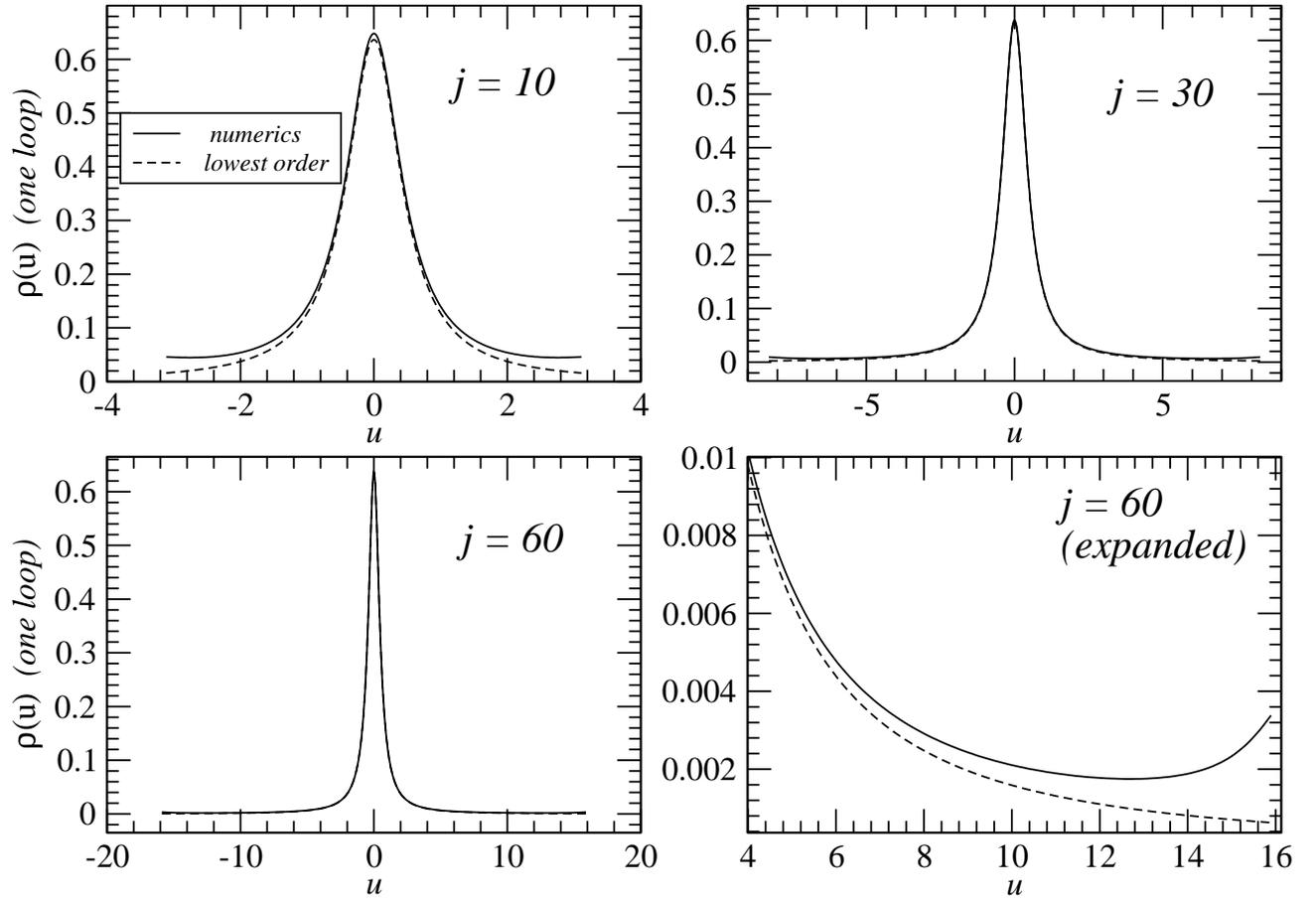
\caption{One loop hole density from the numerical integration of the FRS equation at the two values of the $j$
parameter $j=10, 30$ and $j=60$. In the bottom-right panel, we show a detailed view of the case $j=60$ where one can appreciate
the difference with respect to the lowest order density.}
\label{fig:one-rho}
}

\newpage
\FIGURE{
\vspace*{0.6cm}
\epsfig{file=Figures/oneloop.rhocorr.eps,height=10cm}
\vspace*{0.5cm}
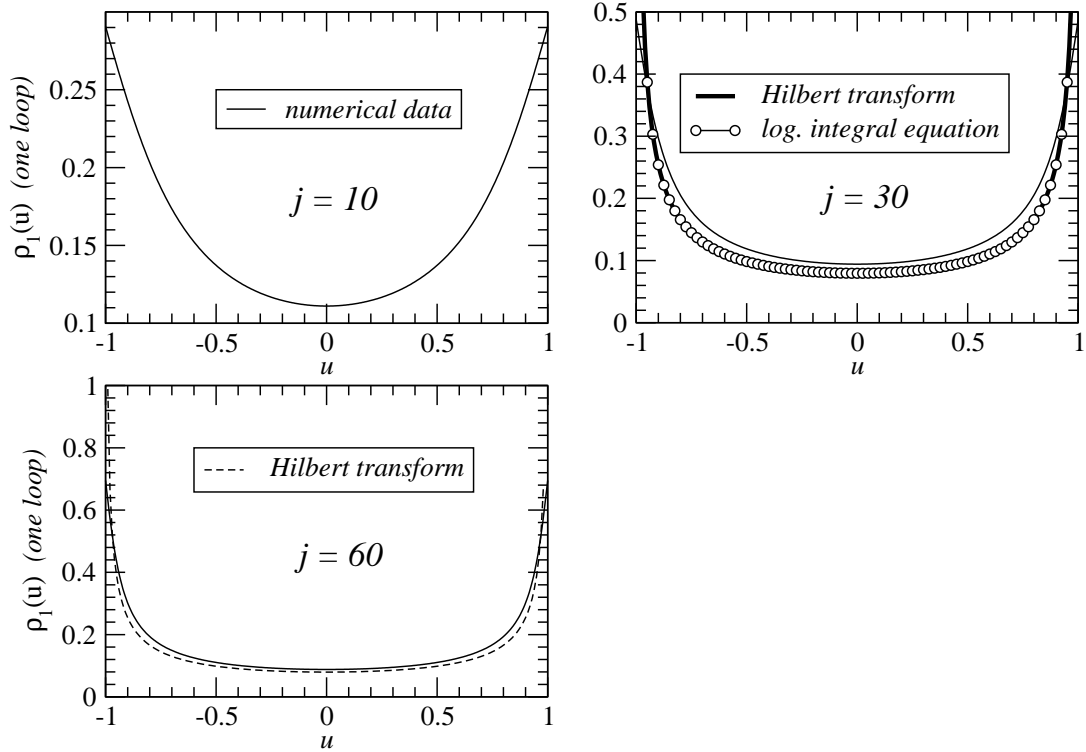
\caption{We show $\rho_1$ which is the one loop hole density from the numerical integration of the FRS equation
minus the analytical lowest order expression of the density. The remaining curves should give at large $j$ the 
NLO correction computed in the text. For $j=30$, we superimpose the analytical expression 
of $\rho_1$ ({\em Hilbert transform} label) and the numerical solution of the logarithmically singular integral equation that
determines it ({\em log. integral equation} label). For $j=60$ we show the numerical data and the {\em Hilbert transform} result.}
\label{fig:one-rho-corr}
}

\newpage
\FIGURE{
\vspace*{0.6cm}
\epsfig{file=Figures/one.gap.eps,height=12cm}
\vspace*{0.5cm}
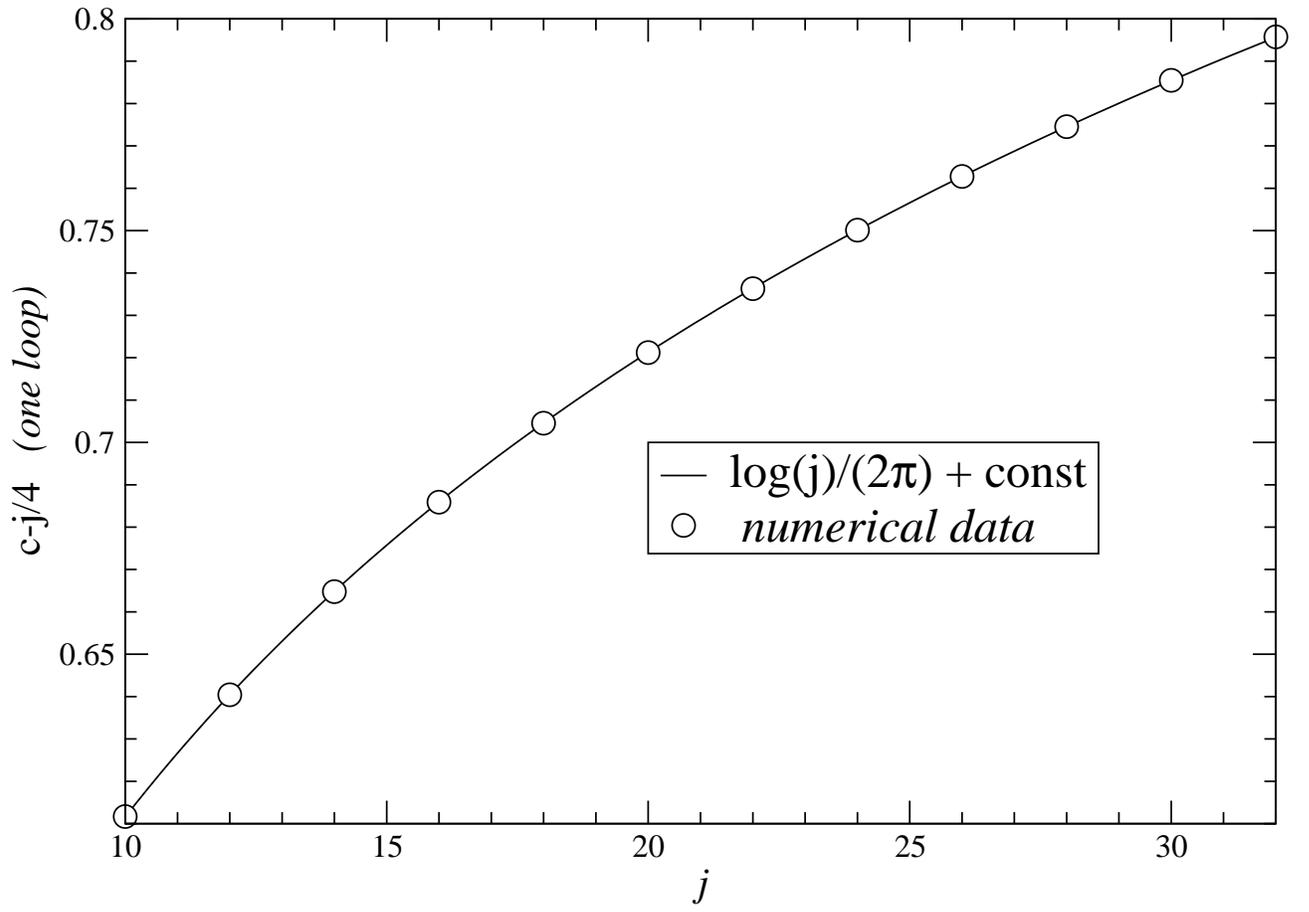
\caption{One loop gap as a function of $j$. We show the numerical data minus the leading order contribution. We also superimpose
a fit with the heuristic logarithmic contribution discussed in the text plus an additional subleading constant.}
\label{fig:one-gap}
}

\newpage
\FIGURE{
\vspace*{0.6cm}
\epsfig{file=Figures/one.gamma.eps,height=12cm}
\vspace*{0.5cm}
\caption{One loop generalized scaling function. We show the numerical data and superimpose two curves with the 
NLO and LO analytical expression.}
\label{fig:one-gamma}
}

\newpage
\FIGURE{
\vspace*{0.6cm}
\epsfig{file=Figures/twoloop.rho.eps,height=12cm}
\vspace*{0.5cm}
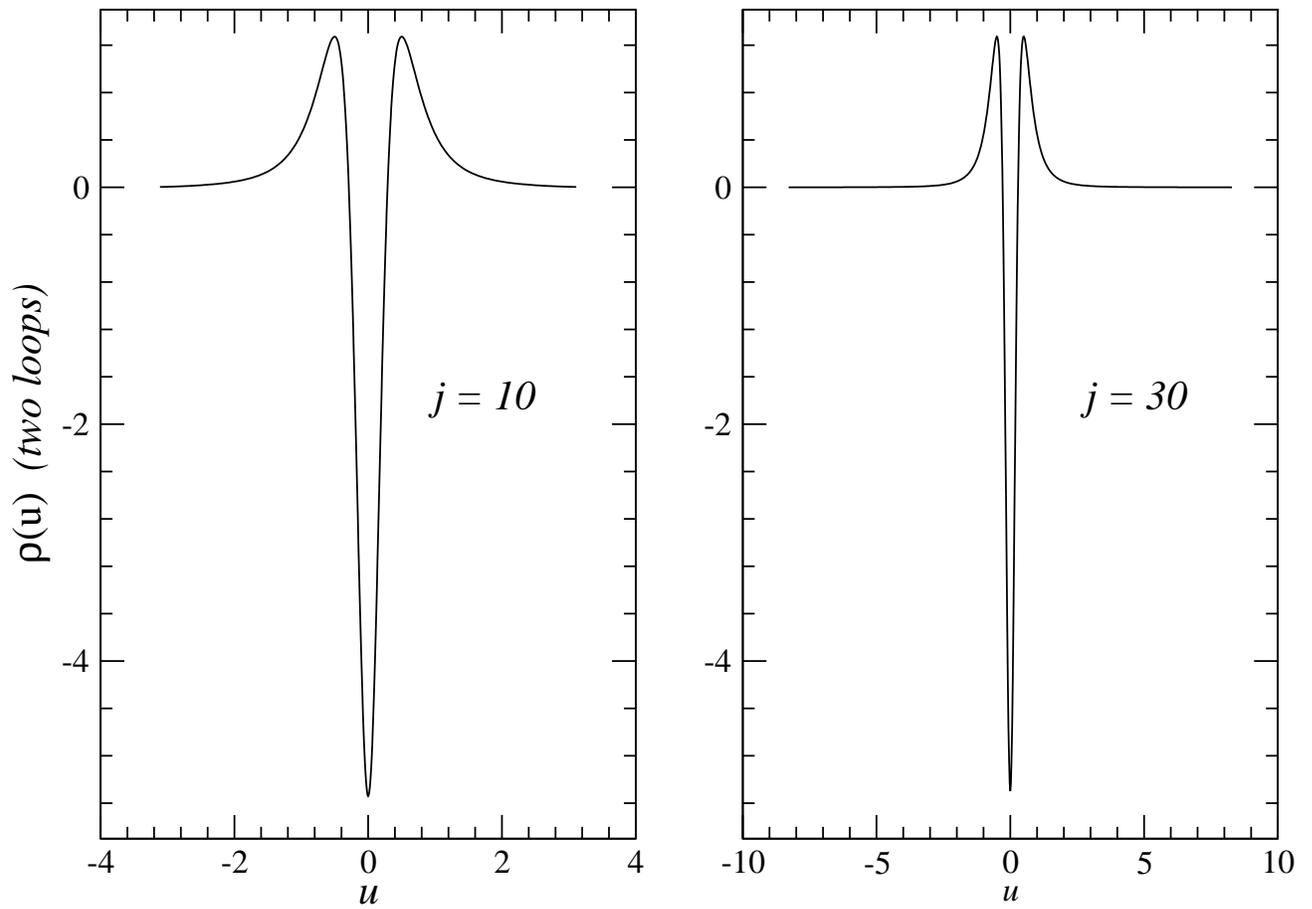
\caption{Two loop hole density from the numerical integration of the FRS equation at the two values of the $j$
parameter $j=10$ and $j=30$. These curves are almost coincident with the leading order density.}
\label{fig:two-rho}
}

\newpage
\FIGURE{
\vspace*{0.6cm}
\epsfig{file=Figures/two.gap.eps,height=12cm}
\vspace*{0.5cm}
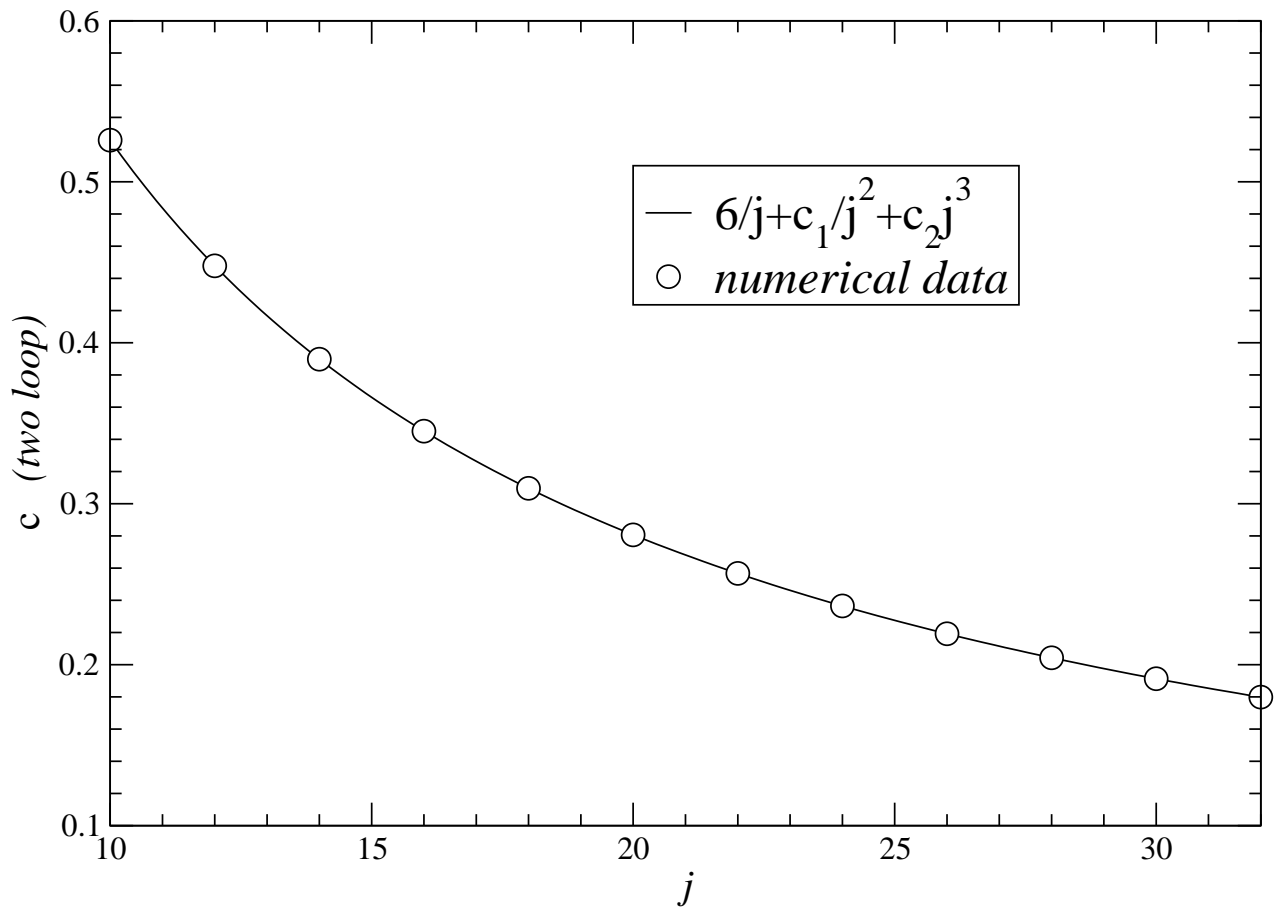
\caption{Two loop gap. We show the numerical data and superimpose a reasonable 2-terms fit including the leading order 
whose coefficient is very accurately reproduced by a would-be 3-terms fit.}
\label{fig:two-gap}
}

\newpage
\FIGURE{
\vspace*{0.6cm}
\epsfig{file=Figures/two.gamma.eps,height=12cm}
\vspace*{0.5cm}
\caption{Two loop generalized scaling function. We show the numerical data and superimpose two curves with the 
NLO and LO analytical expression.}
\label{fig:two-gamma}
}

\end{document}